\documentclass[preprint,12pt]{elsarticle}
\usepackage{graphicx}
\usepackage{amsmath}
\usepackage{amsfonts}
\usepackage{natbib}
\usepackage{dcolumn}
\usepackage{bm}
\usepackage{setspace}

\begin{document}

\begin{frontmatter}

\journal{Journal of Colloid and Interface Science}

\title{A model for colloidal suspension under magnetohydrodynamic conditions}

\author[fys]{Qinghua Chen}
\ead{chenqinghua2002@gmail.com}
\author[cok]{Bernard Stuyven}
\author[fys]{Liviu F. Chibotaru}
\author[cit]{Jan Vermant}
\author[fys]{Johan Vanacken}
\author[cok]{Johan A. Martens}
\author[mtm]{Jan Van Humbeeck}
\author[fys]{Victor V. Moshchalkov}
\ead{victor.moshchalkov@fys.kuleuven.be}

\address[fys]{INPAC-Institute for Nanoscale Physics and Chemistry, K. U. Leuven, Celestijnenlaan 200D,
B-3001 Heverlee, Belgium}
\address[cok]{Center for Surface Chemistry and Catalysis, K. U. Leuven, B-3001 Heverlee, Belgium}
\address[cit]{Applied Rheology and Polymer Processing, K. U. Leuven, B-3001 Heverlee, Belgium}
\address[mtm]{Physical Metallurgy and Materials Engineering, K. U. Leuven, B-3001 Heverlee, Belgium}


\begin{abstract}
We present a comprehensive model to account for the behavior of
suspended nanoparticles under magnetohydrodynamic conditions. The
Lorentz force not only drags nanoparticle flocs toward the walls
reducing the distance between flocs resulting a more negative total
pair interaction potential energy, but also produces extra
magnetic-induced stresses inside a floc leading to a change of pair
interaction distance thus giving rise to a less negative total
potential energy. The model explains quite well the recent
experimental results showing that magnetic field assists aggregation
in laminar or weak turbulent flows, but favors floc disruption in
turbulent regime.
\end{abstract}


\end{frontmatter}


\section{Introduction}
The stability of colloidal suspensions has drawn a lot of interest
in the latest century. According to the well-known DLVO
(Derjaguin-Landau-Verwey-Overbeek) theory
\cite{a.Derjaguin,b.Verwey}, this stability is governed by the
repulsive interaction due to the overlap of electrical double
layers, competing with the Van der Waals attractive interaction
between the primary particles. If the particles are transported by
molecular motion or mechanical mixing, further floc (loose structure
that composed by primary particles and fluid) aggregation occurs due
to the particle collisions \cite{b.Montgomery}. On the other hand,
during the transport process the flocs are subjected to unequal
shearing forces, especially under turbulent conditions, giving rise
to surface erosion  \cite{a.Argaman} or floc disruption
\cite{a.Thomas}. Once the rate of floc breakup becomes equal to the
rate of floc growth the system is in equilibrium and a steady-state
floc size distribution is reached.

Recent decades, magnetic field has been considered by some research
groups as another factor enhancing dispersion. Svoboda \emph{et
al.}\cite{a.Svoboda} and Parker \emph{et al.}\cite{a.Parker} have
mentioned that, in \emph{stationary or gently} flowing suspensions,
magnetic fields magnetize the particles yielding attractive
interaction between magnetic dipoles. The modified DLVO potential
curve produces a secondary minimum at relatively large interparticle
distances, where the colloid becomes unstable and flocculates as
loose aggregates. This magnetic attraction is, however, too small to
be observed. Busch \emph{et al.}\cite{a.Busch} have argued that
combining a \emph{laminar} flow with a magnetic field, \emph{i.e.},
under magnetohydrodynamic (MHD) conditions, a much more pronounced
effect can be obtained experimentally. Based on a hypothesis that
the flowing fluid is conducting, the authors ascribed this effect to
the formation of a boundary layer near the walls leading to a much
stronger velocity gradient to increase the collision frequency and,
thus, to increase the average floc size \cite{a.Busch}.
Unfortunately, this promising concept has not received its matching
attractive case for \emph{turbulent} flow since, to the best of our
knowledge, the existing models cannot explain adequately the
available experimental data. Sometimes these models are even in
\emph{contradiction} with the experiment. Therefore, it is necessary
to develop a convincing model to explain the recent experimental
results, especially the disruption mechanism of flocs under
turbulent MHD conditions.

In this paper we present a comprehensive model to describe the
magnetic effects on a suspension under hydrodynamic conditions. We
assume that a magnetic field produces magnetic (Lorentz) forces
acting on primary-particles and hence flocs, which changes the
distances between the flocs and the distances between
primary-particles inside a floc, thus giving rise to a growth or
reduction of the floc size. Our model predicts that under MHD
conditions, the size increases in laminar or weak turbulent flows
and decreases in turbulent regime, which explain quite well the
experimental data available by now
\cite{a.Busch,a.Bernard,a.Bernard2}.

\section{Models and discussions}
As we know from Refs.~\citenum
{a.Svoboda,a.Parker,a.Busch,a.Bernard,a.Bernard2}, when a transverse
external magnetic field ($B_z$) is applied, the total energy of
interaction of colloidal particles is
\begin{equation}\label{e.totalEnergy}
V_T=V_{EL}+V_{VDW}+V_{M},
\end{equation}
where $V_{EL}=4\pi\varepsilon a\psi_0^2exp[-\kappa a(s-2)]/s$ is the
simple approximate form of the repulsive energy of electrical double
layer, whereby $\varepsilon$ is the absolute dielectric constant of
the medium, $a$ is the radius of interacting spheres, $\psi_0$ is
the potential at the surface of the particles, $\kappa$ is the
Debye-H$\ddot{u}$chel parameter, and $s$ is the ratio of the
distance of the centers of two sphere $R_0$ to there radius $a$;
$V_{VDW}=-\frac{A}{6}[\frac{2}{s^2-4}+\frac{2}{s^2}+ln\frac{s^2-4}{s^2}]$
is the Van der Waals attractive energy between two spherical
particles whereby $A$ is the Hamaker constant; and
$V_M\cong-\frac{32\pi^2a^3\chi^2B_z^2}{9\mu_0s^3}$ is the magnetic
attraction between magnetic dipoles, whereby $\chi$ is the volume
magnetic susceptibility of particles and $\mu_0$ is the magnetic
permeability.

However, under the same magnetic environment, if the suspension is
hydrodynamically transported in a pipe-line system, flocs flowing
along a pipe ($x$-direction) are subjected to Lorentz forces. As it
can be seen from Fig. 1(a), for the flocs with positive charges, the
Lorentz forces
\begin{figure}[tb]
\centering
\includegraphics*[width=8.0cm]{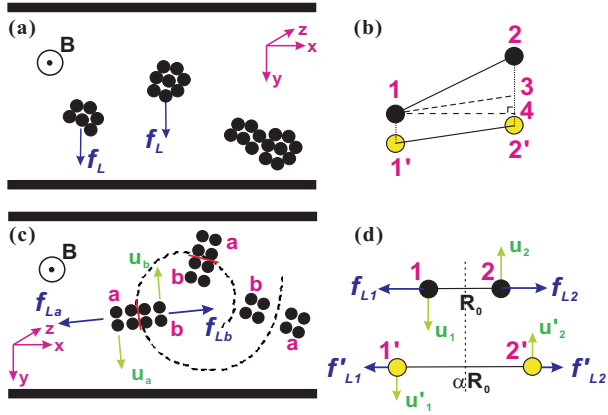}
\caption{\label{fig:migration} (Color online) (a) Migration of flocs
towards the walls in a laminar flow. (b) Schematics of the distance
change between two flocs after magnetic field is applied. Position-
$1$ and $2$ are the positions of the two flocs without a magnetic
field. Position- $1'$ and $2'$ are the positions of the two flocs
with a magnetic field after time $t$. Obviously, distance-$1'2'$ is
somewhat shorter than distance-$12$. (c) Self-rotation of a floc in
a turbulent flow. The dashed line is a vorticity orbit while the
solid line separates the primary-particles into two groups, $a$ and
$b$, which have opposite net vector velocity $u_a$ and $u_b$ with
respect to the applied magnetic field, producing Lorentz forces with
opposite directions $f_{La}$ and $f_{Lb}$. (d) Schematics of the
distance change between two primary-particles in a floc after a
magnetic field is applied. Before disruption, the distance is $R_0$
without a magnetic field but $\alpha R_0$ with a magnetic field. It
is clear that $\alpha\geq1$.}
\end{figure}
always drag the flocs toward the walls. This process is similar to
the sedimentation effect that can be called as \emph{floc-migration
(FM)}.

On the other hand, floc rotation in laminar flows \cite{b.Mason} and
velocity fluctuation in turbulent flows \cite{a.Thomas,b.Heinze} are
inevitable. In most cases, the former one is neglected while the
latter should be taken into account. As mentioned in Ref.~\citenum
{a.Thomas}, random velocity fluctuations may induce pressure
difference on opposite sides of a floc, which causes its deformation
and eventual rupture \cite{a.Thomas}, \emph{i.e.}, it increases the
distance between the primary-particles inside the floc. Thus, as
shown in Fig. 1(c), we can expect that in a turbulent flow a floc
moves along the circular orbit and spins around its center, showing
\emph{floc-self-rotations (FSR)}. This is somewhat similar to the
Earth orbital motion around the Sun combined with the Earth's
spinning around its axis. Such FSR causes the constituting
primary-particles in the floc to have different net vector
velocities. In this case we divide the floc into two parts, $a$ and
$b$, which have opposite velocity directions $u_a$ and $u_b$. When a
magnetic field is applied, the two velocities produce opposite
Lorentz forces and induce an \emph{extra} pressure difference (we
identify it as \emph{magnet-induced stress}) on opposite sides of
the floc, which enhances the total stress (\emph{i.e.} the total
interaction energy $V_T$ between two primary-particles) under
turbulent conditions, and thus stretches the floc to a more prolate
spheroid. Once the total stress surpasses the yield stress of the
floc, it breaks up.

In principle, we cannot directly estimate the FM and FSR effects by
a special potential term as $V_{EL}$, $V_{VDW}$, and $V_{M}$ shown
in Refs.~\citenum {a.Svoboda,a.Parker} because the Lorentz forces
acted on different charged particles are independent from each
other. But, by means of the distance change between the flocs due to
the FM effect, or between the primary-particles inside a floc due to
the FSR effect we can estimate the total energy change in
Eq.~(\ref{e.totalEnergy}). To do this, we need to know the velocity
distribution in each position at every time. The most precise way is
to solve the Navier-Stokes equations by computational fluid dynamics
(CFD) softwares. But this is a quite complicated task and it goes
beyond the scope of this work.

Fortunately, we can have two simple alternative ways to estimate the
velocity distributions and hence the distance changes due to the
magnetic effects. First, for the distance change between flocs
caused by the FM effect, considering that this FM effect is dominant
in laminar flows (also possible in \emph{weak} turbulent flows), one
can approximate the velocity in the $x$-direction along a
2-dimension (2$D$) pipe as
\begin{equation}\label{e.laminar}
v_x=v_{max}(1-4y^2/D^2),
\end{equation}
where $v_{max}$ is the maximum linear velocity at the center part of
the pipe, and $D$ is the diameter of the pipe. Then, from the
physical viewpoint, the dynamics of an \emph{isolated} floc in
$y$-direction is governed by the Newtonian equation as
\footnote{Interactions from Electrical double layer, Van der Waals,
and magnetic dipoles are not taken into account here since they are
balance each other.}
\begin{equation}\label{e.newtonian}
  F_{Gravity,y}+F_{Lorentz,y}+F_{viscosity,y}=m_f\frac{d^2y}{dt^2},
\end{equation}
where $F_{Gravity,y}$ is the gravity force, $F_{Lorentz,y}$ is the
Lorentz force, $F_{viscosity,y}$ is the friction force of the fluid,
and $m_f$ and $y$ are the mass and translation of the floc,
respectively.

If $F_{Lorentz,y}<<F_{Gravity,y}$, the system is reduced to the
sedimentation case and the velocity and moving distance in
$y$-direction, $v_y$ and $y$, are mainly dependent on the gravity
force. Thus, at steady state, one can easily find that
\cite{a.Parker},
$v_y=\frac{dy}{dt}=\frac{2g}{9\mu}(\rho_{particle}-\rho_{fluid})a_f^2$,
and $y=y_0+\frac{2g}{9\mu}(\rho_{particle}-\rho_{fluid})a_f^2t$,
where $\rho_{particle}$ and $\rho_{fluid}$ are the densities of the
floc and the fluid, respectively, $a_f$ is the radius of the floc,
$g$ is the gravity constant, $\mu$ is the viscosity of the fluid,
and $t$ is the integral time.

However, in modern nano-dispersion systems the relation
$F_{Lorentz,y}>>F_{Gravity,y}$ is often satisfied since the gravity
effect can be neglected. Similarly, by using Eqs.~(\ref{e.laminar})
and (\ref{e.newtonian}) the $v_y$ and $y$ can be estimated as,
\begin{subequations}
\label{e.migration}
\begin{gather}
v_y=\frac{dy}{dt}=\frac{AD[-1+tanh(At)^2](D^2-4y_0^2)}{2[D-2y_0tanh(At)]^2},\label{e.migration1}\\
y=\frac{2y_0D-D^2tanh(At)}{2D-4y_0 tanh(At)},\label{e.migration2}
\end{gather}
\end{subequations}
where $A=qNB_zv_{max}/(3\pi a_f\mu D)$, $q$ is the elementary charge
constant, and $N$ is the total charge number in the floc which is
associated with the potential of the surface charges, $\psi_0$.

Consequently, combining Eq.~(\ref{e.laminar}) and
(\ref{e.migration}) we can estimate the position of a floc flowing
in a laminar flow at any time due to the FM effect. Fig. 1(b) is the
schematic presentation of two flocs migrating in a 2$D$ pipe.
Position- $1$ and $2$ are the originals of the two flocs. As it can
be seen, without magnetic field the $y$ coordinates are not changed,
while with magnetic field the flocs move to position- $1'$ and $2'$,
respectively, after time $t$. Obviously, distance-$1'2'$ is somewhat
shorter than distance-$12$. To trace in a better way the distance
change by the magnetic field we introduce a ratio
$\alpha=\frac{distance-1'2'}{distance-12}$. This $\alpha$ is clearly
dependent on magnetic field, and potential of the surface charge.
For such FM case $\alpha$ is smaller than 1.

Secondly, the distance change between the primary-particles inside a
floc, due to the FSR effect which dominates in turbulent flows, can
also be estimated in a simple way. \emph{Notice that before a floc
breaks the extra stress (force), for instance, the magnetic-induced
stress inside the floc is always relaxed by the distance change.}
For simplicity we only discuss pair interaction between any two
primary-particles as the classic DLVO theory did to estimate the
distance change between the pairs. Fig. 1(d) is the schematic
presentation of the distance change between two primary-particles
inside a floc induced by magnetic field. We assume that these two
primary-particles are balance at positions $1$ and $2$ without
magnetic field and at positions $1'$ and $2'$ with magnetic field.
The distances between them are $R_0$ and $\alpha R_0$, respectively.
For a given turbulent dispersion system without magnetic field, if
we assume that the length and time scales are the order of
Kolmogorov length and time scales, respectively, the distribution of
the intensity of turbulence over the range of wavenumbers (or over
the range of eddy sizes) can be determined as follow
\cite{b.Heinze},
\begin{equation}\label{eq.intensity}
   \frac{3}{2}u^2_{1,2}=\int_k^\infty E(k,t)dk,
\end{equation}
where $u_{1,2}$ is the fluctuation velocity, $k$ is the wavenumber
(reciprocal of the characteristic length $r$), and
$E(k,t)=const.\times \epsilon^{2/3}k^{-5/3}$ for $\eta\ll r \ll L$
while $\eta$ is the Kolmogorov length scale and $L$ is the large
scales. Therefore, before disruption the distance between any two
primary-particles, $R_0$, inside a floc is governed by the force
balance equation acting at one primary-particle as,
\begin{equation}\label{e.forceBalance0}
  \left.\frac{\frac{1}{2}m_{1,2}u_{1,2}^2}{R_0/2}=-\frac{dV_T}{ds}\right|_{s=R_0/a},
\end{equation}
where $m_{1,2}$ is the mass of primary-particles. The left-hand side
is the force caused by the kinetic energy of the turbulence while
the right-hand side is the static force caused by the total
interaction potential. Similarly, for a system with a magnetic
field, the force balance is changed to,
\begin{equation}\label{e.forceBalance1}
  \left.\frac{\frac{1}{2}m_{1,2}u_{1,2}'^2}{\alpha R_0/2}+qN_0u'_{1,2}B_z=-\frac{dV_T}{ds}\right|_{s=\alpha R_0/a},
\end{equation}
where $u'_{1,2}$ is the fluctuation velocity with a magnetic field,
and $N_0$ is the total charge number in the primary- particle. The
second term in the left-hand side is the Lorentz force. Furthermore,
before disruption, the change of the total interaction potential
must equal to the change of the kinetic energy (energy
conservation),
\begin{equation}\label{e.energyConservation}
  V_T(\alpha R_0)-V_T(R_0)=\frac{1}{2}m_{1}(u_1'^2-u_1^2)+\frac{1}{2}m_{2}(u_2'^2-u_2^2).
\end{equation}
For simplicity, we further suppose that the primary-particles are
identical, that is, $m_1=m_2=m_{pp}$, $|u_1|=|u_2|=|u|$ and
$|u'_1|=|u'_2|=|u'|$. Thus, combining Eqs.~(\ref{e.totalEnergy}),
(\ref{e.forceBalance0}), (\ref{e.forceBalance1}), and
(\ref{e.energyConservation}) we can calculate the four unknown
variables, $V_T$, $R_0$, $u'$, and $\alpha$. Obviously, this
$\alpha$ is dependent also on the magnetic field, Reynolds number,
and potential of the surface charge. In this case $\alpha$ is
greater than 1, which means the floc is getting more prolate and
less oscillating before it is broken apart.

Now we turn to the discussion of the effect of an external
transverse magnetic field on the pair interaction potential curve.
Fig. 2 presents the total energy with respect to the distance
between two primary-particles. The solid line without symbol is the
classic DLVO curve without magnetic field.
\begin{figure}[tb]
\centering
\includegraphics*[width=8.0cm]{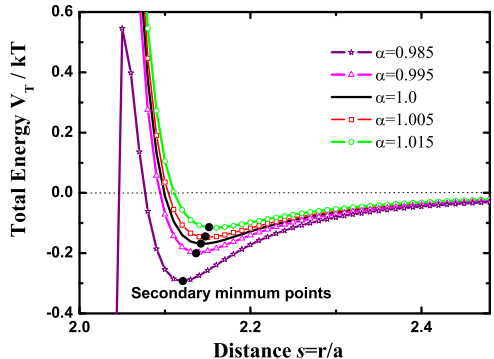}
\caption{\label{fig:DLVO} (Color online) Total energy with respect
to the distance between two primary-particles. The solid black line
without symbols is the classic DLVO curve without magnetic field.
Solid curves with open star, triangle, square, and circle are the
cases of $\alpha=0.985, 0.995, 1.005, and~1.015$, respectively.
Black filled dots in each curve are the secondary minimum points.}
\end{figure}
Solid curves with open star, triangle, square, and circle are slight
distance change cases, such as, $\alpha=0.985, 0.995, 1.005,
and~1.015$, respectively. From the figure we can find that in
laminar flow ($\alpha<1$), due to the FM effect, the curve produces
a more pronounced secondary minimum which means that the interaction
energy is more attractive than the one without magnetic field. In
this case the flocs flocculate more easily from a relatively large
interparticle distance, giving rise to a relatively large floc size,
which is in a good agreement with the experimental results reported
in Refs.~\citenum {a.Busch,a.Bernard,a.Bernard2}. While in a
turbulent flow ($\alpha>1$), due to the FSR effect, the curve is
completely above the classic one which means that the total energy
is less attractive than the one without magnetic field. In this case
the original aggregates are elongated into prolate spheroids by the
extra magnetic-induced stress and eventually break up into smaller
flocs leading to a relatively small average floc size. This is the
reason why the size of flocs is reduced in strong turbulent flows
reported in Refs.~\citenum {a.Bernard,a.Bernard2}. Furthermore, in
weak turbulent flows, both the FM and FSR effects are comparable.
Thus, the floc size is dependent on the competition between FM and
FSR. Using this model we can explain why the floc size increases in
weak turbulent region in Refs.~\citenum {a.Bernard,a.Bernard2}: the
FM effect in this case is stronger than the FSR.

\section{Conclusions}
To conclude, a transverse magnetic field results in Lorentz forces
acting on primary-particles and hence flocs, which drag flocs toward
the walls and/or elongate flocs into prolate spheroids. The former
mechanism mainly works in laminar flows leading to a decrease of the
distances between the flocs, while the latter dominates in turbulent
flows thus giving rise to an increase of the distances between the
primary-particles inside a floc. This model predicts that under MHD
conditions, the size increases in laminar or gentle turbulent flows
and decreases in turbulent regime. We believe that the presented
model is not only applicable for physics and chemical engineering
science, but also for the biology, medicine, \emph{etc}. A good
example here is the investigation of the stability of protein
molecules. It should be noted here that the model in this paper is
only one eddy for the regime of turbulence. Further stage
calculations one should take into account many eddies of different
sizes and a statistical description should be used.

\section{Acknowledgement}
This work was sponsored by K.U. Leuven Interdisciplinary research programme (IDO-project), IDO/04/009.

\end{document}